\def\year{2021}\relax
\documentclass[letterpaper]{article} 
\usepackage{adjustbox}
\usepackage{makecell}
\usepackage{aaai21}  
\usepackage{times}  
\usepackage{helvet} 
\usepackage{courier}  
\usepackage[hyphens]{url}  
\usepackage{graphicx} 
\usepackage{natbib}  
\usepackage{caption} 
\nocopyright

\title{VaccinItaly: Monitoring Italian Conversations Around Vaccines On Twitter And Facebook}
\author{Francesco Pierri\textsuperscript{1,*}, Andrea Tocchetti\textsuperscript{1}, Lorenzo Corti\textsuperscript{1}, Marco Di Giovanni\textsuperscript{1,2},\\  Silvio Pavanetto\textsuperscript{1}, Marco Brambilla\textsuperscript{1}, Stefano Ceri\textsuperscript{1}\\
}
\affiliations{\textsuperscript{1}
Dipartimento di Elettronica, Informazione e Bioingegneria\\ Politecnico di Milano, Milano, Italy\\
\textsuperscript{2} Università di Bologna, Bologna, Italy\\
\textsuperscript{*}Corresponding author. E-mail: francesco.pierri@polimi.it
}
\begin{document}

\maketitle

\begin{abstract}
We present VaccinItaly, a project which monitors Italian online conversations around vaccines, on Twitter and Facebook. We describe the ongoing data collection, which follows the SARS-CoV-2 vaccination campaign roll-out in Italy and we provide public access to the data collected. We show results from a preliminary analysis of the spread of low- and high-credibility news shared alongside vaccine-related conversations on both social media platforms. We also investigate the content of most popular YouTube videos and encounter several cases of harmful and misleading content about vaccines. Finally, we geolocate Twitter users who discuss vaccines and correlate their activity with open data statistics on vaccine uptake. We make up-to-date results available to the public through an interactive online dashboard associated with the project. The goal of our project is to gain further understanding of the interplay between the public discourse on online social media and the dynamics of vaccine uptake in the real world.
\end{abstract}

\section{Introduction}
On January 30th, 2020, the World Health Organization declared the outbreak of a novel coronavirus (SARS-CoV-2) a global pandemic\footnote{\url{https://www.who.int/publications/m/item/covid-19-public-health-emergency-of-international-concern-(pheic)-global-research-and-innovation-forum}}. A year later, the spread of the virus has caused over 121 $M$ confirmed cases and more than 2.5 $M$ fatalities globally\footnote{\url{https://covid19.who.int}}. Italy, in particular, has been one of the first European countries to be hit by the virus, with over 3.28 $M$ confirmed cases and 100 $k$ fatalities as of March 2021, and the first country outside China to implement national lockdown to circumvent its spreading with severe social and economic consequences \cite{bonaccorsi2020, spelta2020after}.
Despite the global crisis, we witnessed the most rapid vaccine development for a pandemic in history when the Pfizer-Biontech vaccine showed a 95\% efficacy and was approved in several countries\footnote{\url{https://www.pfizer.com/news/press-release/press-release-detail/pfizer-and-biontech-conclude-phase-3-study-covid-19-vaccine}} in late Fall, 2020. In the next few months, several other vaccines were going to be approved and made available to the public\footnote{\url{https://www.nytimes.com/interactive/2020/science/coronavirus-vaccine-tracker.html}}. Italy, specifically, has started its vaccination campaign on December 27th, 2020, and reached over 6 $M$ dispensed doses\footnote{\url{http://www.salute.gov.it/portale/nuovocoronavirus/dettaglioContenutiNuovoCoronavirus}} as of March 13th, 2021.

As COVID-19 was spreading around the world, online social networks experienced a so-called "infodemic", i.e. an over-abundance of information about the ongoing pandemic, which yield severe repercussions on public health and safety \cite{zarocostas2020fight, yang2020covid,gallotti2020assessing,guarino2021information}. It is believed that low-credibility information might drive vaccine hesitancy and make it hard to reach herd immunity \cite{yang2020covid, pierri2021impact}. The European Social Observatory for Disinformation and Social Media Analysis has recently identified four macro-categories of unreliable information about COVID-19 vaccines\footnote{\url{https://www.disinfobservatory.org/disinformation-about-covid-19-and-vaccines-a-journey-across-europe/}}: (1) there haven't been enough tests on vaccines to guarantee their safety; (2) causal association for individuals who died after being vaccinated; (3) there are further medical complications due to vaccines; (4) vaccines can modify our DNA.

Since the 2016 US presidential elections, the research community has mostly focused its attention on political disinformation and election-related manipulation of online conversations \cite{Lazer-fake-news-2018,shao2018spread,ferrara2016rise,Pierri2019}. However, much concern has grown around health-related misinformation which became manifest during recent measles outbreaks \cite{filia2017ongoing} and other epidemics such as H1N1 and Ebola \cite{chew2010pandemics,fung2014ebola}, eroding public trust in governments and institutions and undermining public countermeasures during such crises \cite{scotti2020online,digiovanni}.


In this paper, we describe VaccinItaly, a project to monitor Italian conversations around vaccines on multiple social media (Twitter, Facebook) with the aim of understanding the interplay between online public discourse and the vaccine roll-out campaign in Italy. Using a set of Italian vaccine-related keywords, regularly updated to capture trending hashtags and relevant events, as of March 13th, 2021 we collected over 3 M tweets and 1 M Facebook posts published by public pages and groups (we started our collection on December 20th, 2020). We provide public access to the list of keywords and tweet IDs\footnote{\url{https://github.com/frapierri/VaccinItaly}}, whereas access to Facebook data is granted by Crowdtangle\cite{crowdtangle} to academics and researchers upon request\footnote{Nevertheless, we provide a script to replicate our data collection using Crowdtangle keys.}.

A specific goal of our project is to investigate the spread of reliable and unreliable information related to vaccines. Following a huge corpus of literature \cite{Lazer-fake-news-2018,Bovet2019,Grinberg,deverna2021covaxxy,yang2020covid}, we use a consolidated source-based approach to study how news articles, originated from low- and high-credibility websites (see next sections for more details), are shared alongside vaccine-related conversations on the two platforms. We also highlight YouTube as an additional potential source of misinformation about vaccines. Finally, we geolocate over 1 M users on Twitter and correlate their online activity with open data statistics about the Italian vaccine roll-out campaign\footnote{The data are available here: \url{https://github.com/italia/covid19-opendata-vaccini}}.

\begin{figure}[!t]
    \centering
    \includegraphics[width=\linewidth]{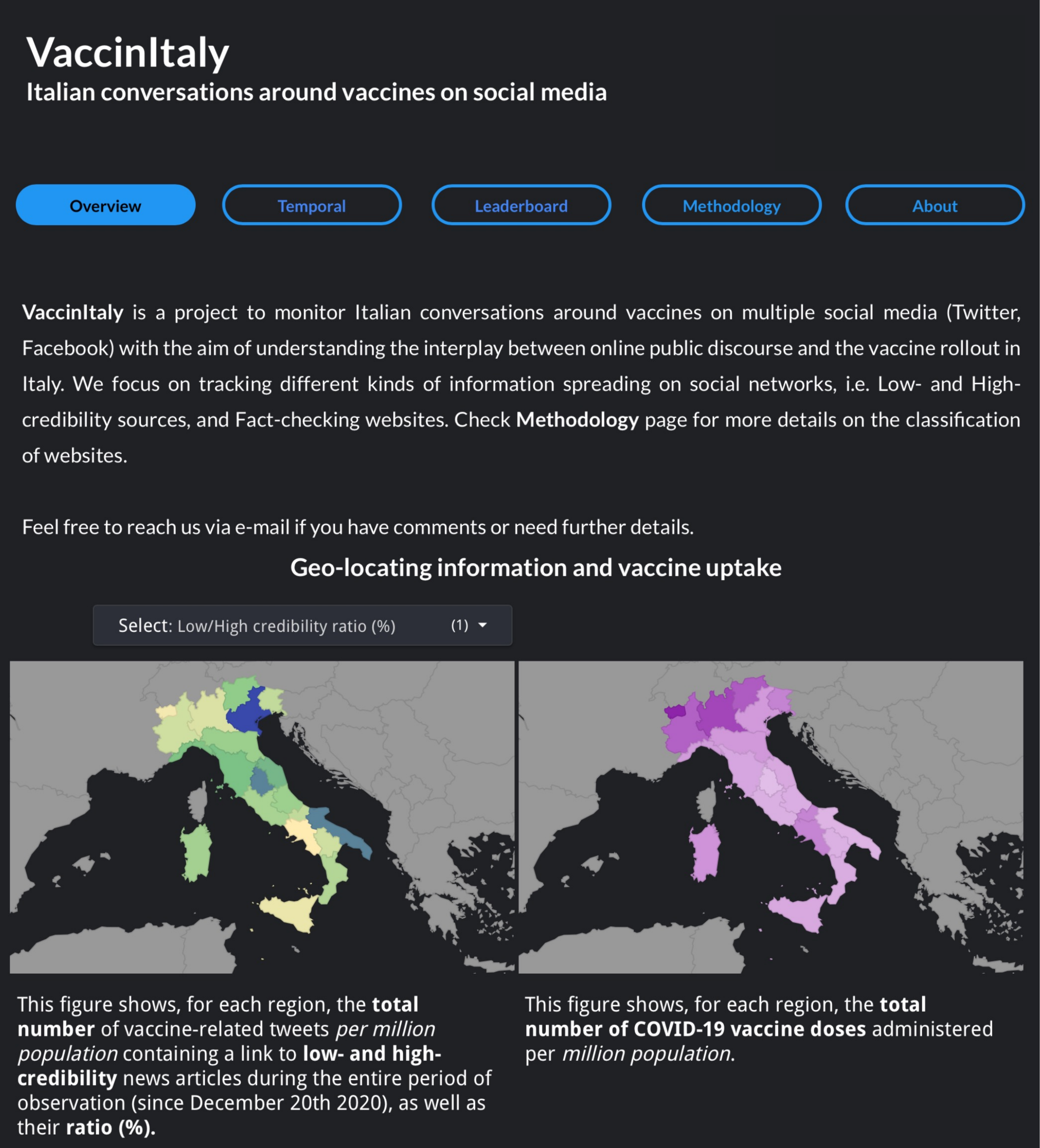}
    \caption{Screenshot of the online dashboard associated to our project. Users can navigate through several sections, each providing different kind of analyses. }
    \label{fig:dashboard}
\end{figure}

Up-to-date results from our ongoing analyses are also available to the public through an online dashboard accessible here: \url{http://genomic.elet.polimi.it/vaccinitaly/}. A preview of the dashboard is available in Figure \ref{fig:dashboard}. This is similar in spirit to CoVaxxy\footnote{\url{https://osome.iu.edu/tools/covaxxy}}\cite{deverna2021covaxxy}, a project based at the Observatory of Social Media (Indiana University) which aims to show the interplay between English-language online misinformation on Twitter and the US vaccine roll-out campaign. However, we focus on the Italian scenario and we also analyze Facebook data.

We believe that our project can contribute to a deeper understanding of the impact of online social networks in an unprecedented scenario where trust in science and governments will be critical to battle a global pandemic.


\section{Related work}
There is a huge corpus of literature around the diffusion of health-related (dis)information on online social networks. We describe a few contributions which are related to the Italian context and refer the reader to \cite{wang2019systematic} for a deeper review of the existing literature on the subject.

\citet{aquino2017web} explored the relationships between Measles, mumps, and rubella (MMR) vaccination coverage in Italy and online search trends and social network activity from 2010 to 2015. Using a set of keywords related to the controversial link between MMR vaccines and autism, originated from a discredited 1998 paper, authors analyzed Google (search) Trends as well as the activity of Facebook pages and Twitter users on the same subject. They reported a significant negative correlation with the evolution of vaccination coverage in Italy (which decreased from 90\% to 85\% during the period of observation). They also identified real-world triggering events which most likely drove vaccine hesitancy, i.e. Court of Justice sentences that ruled in favor of a possible link between MMR vaccine and autism.

\citet{donzelli2018misinformation} provide a quantitative analysis of the Italian videos published on YouTube, from 2007 to 2017, about the link between vaccines and autism or other serious side effects in children. They showed that videos with a negative tone were more prevalent and got more views than those with a positive attitude. However, they did not inspect how videos were treating the link between vaccines and autism.

\cite{righetti_2020} analyzed the Italian vaccine-related environment on Twitter in correspondence with the child vaccination mandatory law promulgated in 2017. Using a keyword-based data collection similar to ours, the author showed that the strong "politicization" of the debate was associated with an increase in the amount of problematic information, such as conspiracy theories, anti-vax narratives, and false news, shared by online users.

\citet{cossard_2020} also analyzed the debate about vaccinations in Italy on Twitter, following the mandatory law promulgated in 2017. They inspected the network of interactions between users, and they identified two main communities of people classified as "vaccine advocates" and "vaccine skeptics", in which they find evidence of echo chamber effects. Besides, they proposed a methodology to predicting the community in which a neutral user would fall, based on a content-based analysis of the tweets shared by users in the two groups.

\section{Data collection}
\subsection{Twitter}
Starting on December 20th, 2020, we use Twitter Filter\footnote{\url{https://developer.twitter.com/en/docs/twitter-api/v1/tweets/filter-realtime/api-reference/post-statuses-filter}} API to collect tweets matching the set of keywords in Table \ref{tab:keywords}, in real-time. We routinely check for trending hashtags and relevant events to add new peculiar keywords, e.g. "\#novaccinoainovax" and "\#iononsonounacavia" were hashtags trending on specific days and consequently they were added to the list of keywords. The latter refers to vaccine advocates stating that no-vax should not be vaccinated, and the former indicates vaccine skeptics who "do not want to be guinea pigs for vaccines". The overall data up to March 13th, 2021 comprises approximately 3 $M$ tweets shared by 258 $k$ unique users.

\begin{table}[!t]
\centering
\resizebox{\linewidth}{!}{%
\begin{tabular}{|l|l|l|}
\hline
vaccini & vaccinarsi & vaccinerai \\ \hline
vaccino & vaccinare & vaccineremo \\ \hline
vaccinazioni & vacciniamoci & vaccinerete \\ \hline
iononmivaccino & vaccinareh24 & iononmivaccinero \\ \hline
vaccinazione & vaccinerò & novaccinoainovax \\ \hline
vaccinocovid & vaccinoanticovid & iononsonounacavia \\ \hline
\end{tabular}%
}
\caption{List of keywords used to filter relevant tweets and Facebook posts. They all refer to vaccines and vaccination in general, and some indicate specific pro and anti-vax views (e.g. "iononmivaccino" means "I will not get vaccinated", "vaccinareh24" means "Vaccinate all day long").}
\label{tab:keywords}
\end{table}

\subsection{Facebook}
We used the \textit{posts/search} endpoint of the CrowdTangle API \citep{crowdtangle} to collect public posts shared by pages and groups which matched the list of keywords previously defined, resulting in over 10 $M$ posts published by over 60 $k$ public pages and groups, and re-shared over 100 $M$ times, as of March 13th, 2021. In the following, we will use the number of shares to compare Facebook with Twitter.

A limitation to our collection of Facebook is the coverage of pages and groups, whose data can be retrieved using the API. The tool includes over 6M Facebook pages and groups: all those with at least 100k followers/members and a very small subset of verified profiles that can be followed like public pages. Besides, some pages and groups with fewer followers and members can be included by CrowdTangle upon request from users. This might bias the data as, for instance, researchers and journalists might be interested in monitoring pages and groups sharing low-credibility thus leading to an over-representation of such content.

\subsection{Sources of low- and high-credibility information}
We extract URLs contained in tweets and Facebook posts to understand the prevalence of low- and high-credibility information shared in vaccine-based conversations \cite{yang2020covid}. We use a consolidated source-based approach to label news articles \cite{Lazer-fake-news-2018,PierriArtoni2020,gallotti2020assessing,shao2018spread,Pierri2020epj,Grinberg,deverna2021covaxxy,Pierri2020scirep,PierriWWW20} depending on the reliability of the source, referring to two lists of Italian low- and high credibility news websites. The former corresponds to websites flagged by Italian fact-checkers for publishing false news, hoaxes and conspiracy theories\footnote{See \url{www.pagellapolitica.it}, \url{www.facta.news} and \url{www.butac.it}}); the latter corresponds to Italian traditional and most popular news websites\cite{vicario2017}, and it is used as a reference to understand the prevalence of misleading and (potentially) harmful information.
Lists are available in our repository\footnote{\url{https://github.com/frapierri/VaccinItaly}}, and we plan to manually augment them during our analyses.


We are aware that this approach, widely adopted in the research community, is not 100\% accurate, as cases of misinformation on mainstream websites are not rare and, similarly, low-credibility websites do not publish solely "fake news". However, to date, it is the most reliable and scalable way to study misleading and harmful information. Another limitation to our estimates is that our lists might not fully capture the amount of low- and high-credibility information circulating on Twitter. Besides, we do not consider different typologies of content such as photos, videos, memes, etc.

\begin{figure}[!t]
    \centering
    \includegraphics[width=\linewidth]{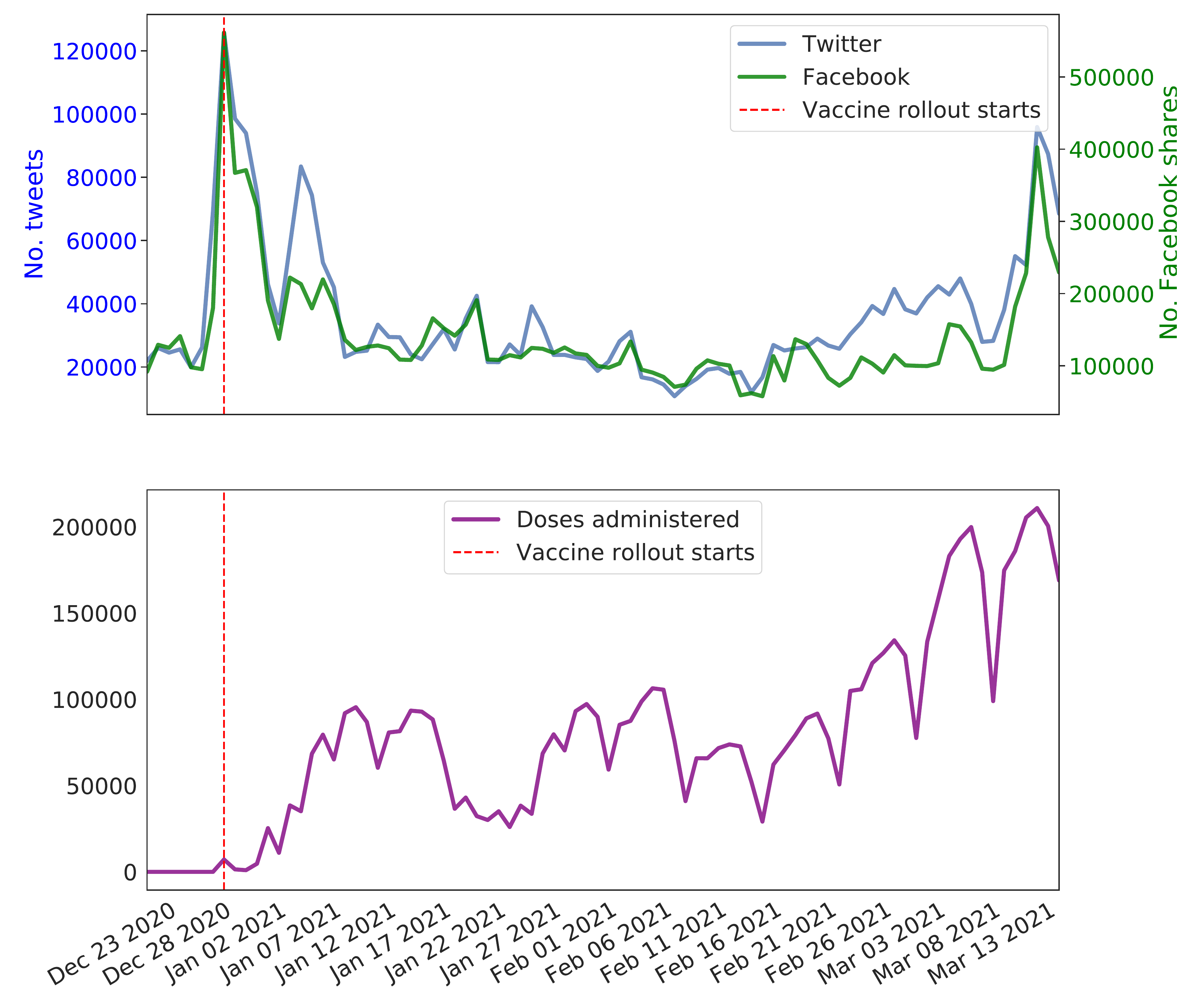}
    \caption{\textbf{Top.} Temporal evolution of the daily volume of vaccines-related posts shared on both Twitter and Facebook. We use a dashed red line to indicate the beginning of the Italian vaccination campaign (27th December, 2020). \textbf{Bottom.} Total number of vaccine doses administered over time.}
    \label{fig:overview}
\end{figure}

\section{Online conversations and vaccine roll-out campaign}
As previously mentioned, we started our collection on December 20th, 2020, in order to capture the beginning of the Italian vaccination campaign. A symbolic start took place on December 27th 2020, when a few thousand doses of Pfizer–BioNTech COVID-19 vaccine were used to vaccinate part of the medical and health personnel of hospitals, while a few days after 2021 New Year's eve over 300 $k$ doses were delivered to Italy. In this ongoing phase, the priority is given first to health medical and administrative personnel, together with the guests and personnel of nursing homes, and then to elderly people and public service personnel.

Accordingly, we notice a huge spike in both Twitter and Facebook volumes following the symbolic start (over 120 $k$ tweets and 500 $k$ Facebook posts shared in a single day), and a slightly smaller spike after the actual beginning of the campaign (a peak of 80 $k$ tweets and 400 $k$ Facebook shares), as shown in Figure \ref{fig:overview}. As the number of doses administered increases to a steady level, we notice that public attention slowly decreases. However, we notice a second surge of online conversations in March, in correspondence with the suspension of Astrazeneca vaccine in several European countries following an investigation of the European Medicines Agency about unusual blood disorders\footnote{\url{https://www.ema.europa.eu/en/news/covid-19-vaccine-astrazeneca-prac-preliminary-view-suggests-no-specific-issue-batch-used-austria}}.

Overall we notice that the volume of vaccine-related conversations on Facebook is much higher than on Twitter, and this is probably due to the different size of their user base\footnote{\url{https://www.statista.com/statistics/787390/main-social-networks-users-italy/}}.

\begin{figure}[!t]
    \centering
    \includegraphics[width=\linewidth]{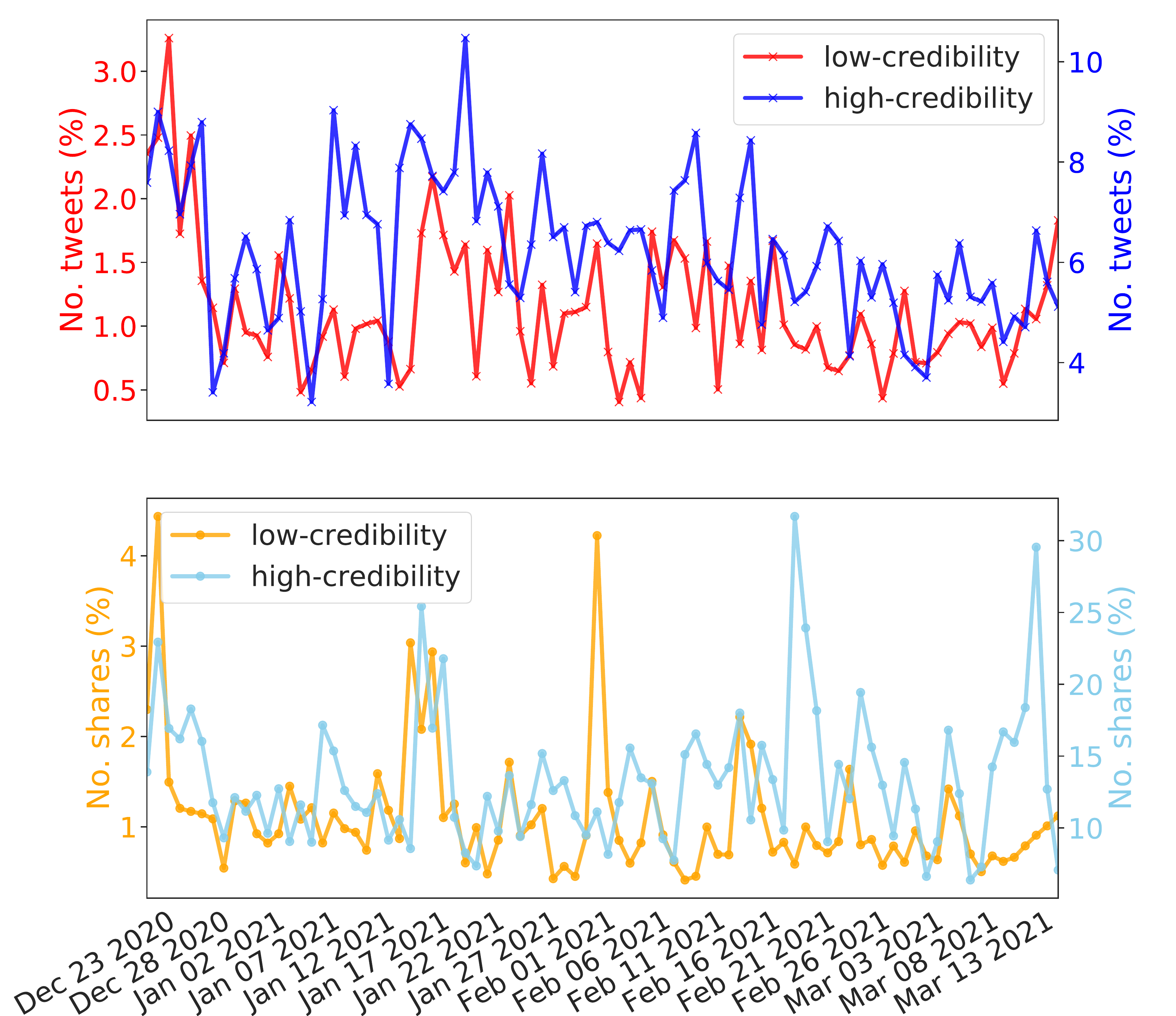}
    \caption{Daily fraction  of high-credibility and low-credibility content, compared to online conversations altogether, for Twitter (Top) and Facebook (Bottom).
    }
    \label{fig:news_ts}
\end{figure}

\section{Prevalence of low- and high-credibility information}
A key focus of our research is to analyze the spread of low-credibility information on social media, using high-credibility information as a reference. Overall, we report over 30 $k$ tweets and 130 $k$ Facebook shares linking to low-credibility news, and over 188 $k$ tweets and 1.6 $M$ Facebook shares linking to high-credibility news.


In Figure \ref{fig:news_ts}, we plot the fraction of tweets and Facebook posts shared that contain a link to either low- or high-credibility information. We note that the amount of low-credibility articles shared on both social media is much smaller compared to high-credibility, on both platforms. Relatively, the mean daily amount of low-credibility information is similar on the two platforms (1.13\% on Twitter, 1.10\% on Facebook), whereas, interestingly, the mean daily amount of high-credibility circulating on Facebook is higher compared to Twitter (6.27\% on Twitter, 13.41\% on Facebook). Given the limitations of our analysis, we might not simply state that the information spreading on Facebook is more reliable than on Twitter. Besides, the amount of low-credibility information is non-negligible on both platforms, and it might play a relevant role in shaping the public discourse and opinion around vaccines.

\begin{figure*}[!ht]
    \centering
    \includegraphics[width=0.8\linewidth]{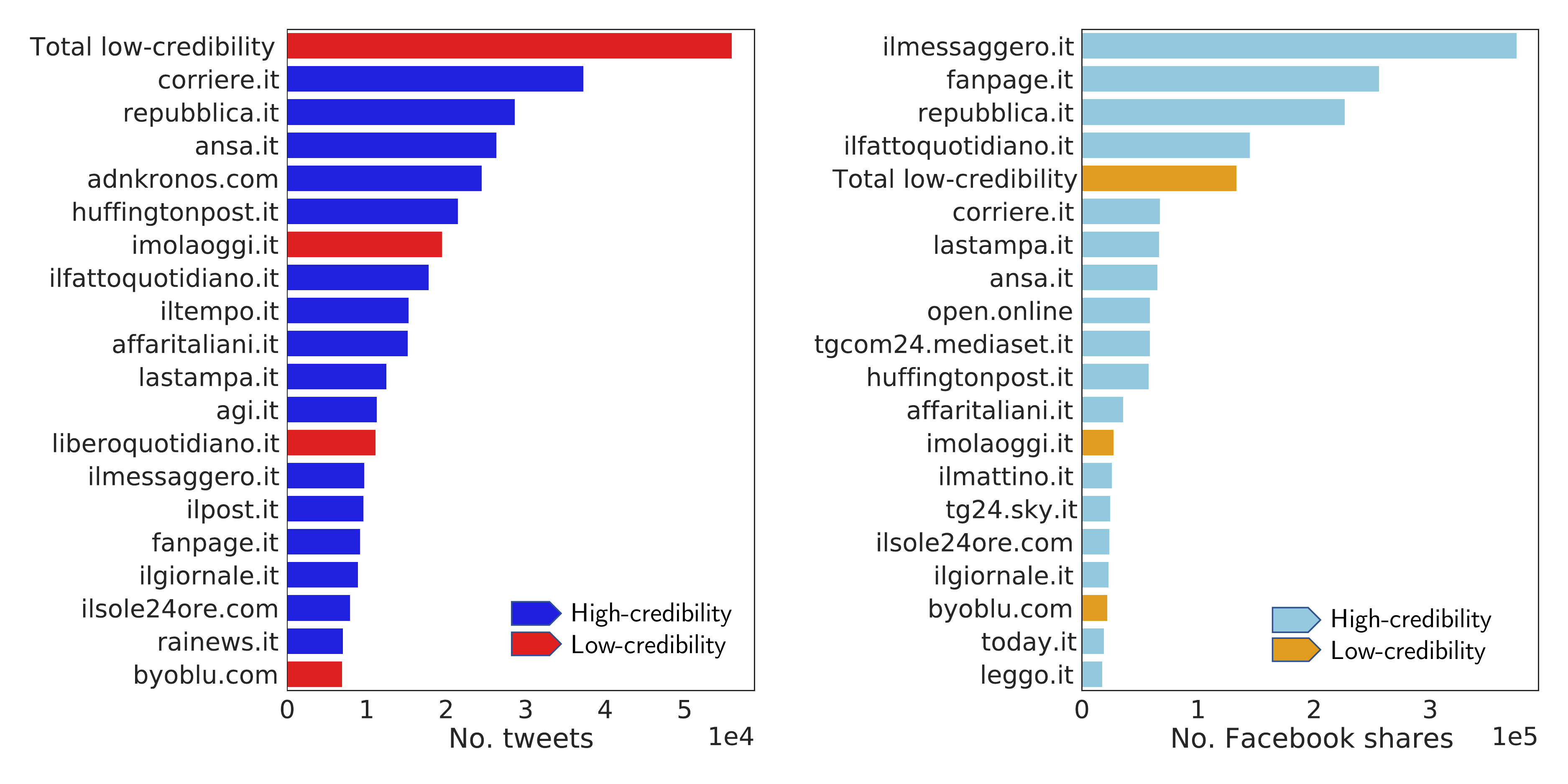}
    \caption{Top-20 news w.r.t the overall number of Twitter (left) and Facebook (right) shares. We also indicate the total amount of low-credibility content and compare it with individual sources of high-credibility information.}
    \label{fig:news_leaderboard}
\end{figure*}

In Figure \ref{fig:news_leaderboard}, we show a leaderboard of the top-20 news sources shared on Facebook and Twitter, considering both low- and high-credibility information. We also add the totality of low-credibility information. As previously noted, Facebook shares are an order of magnitude larger than Twitter.
Besides, high-credibility domains are shared more than low-credibility websites on both platforms. Except for "liberoquotidiano.it", a right-wing news website which notably publishes misleading information, we notice the same two most shared low-credibility domains in the leaderboard, namely "imolaoggi.it" and "byoblu.it". The former is a well-known far-right-wing website that regularly publishes false news with nationalist and anti-immigration views, the latter is a blog that has been repetitiously flagged for sharing hoaxes about health-related subjects, including the COVID-19 pandemic.

By looking at the overall amount of low-credibility news shared on both platforms, we notice interestingly that on Twitter this is larger than any individual high-credibility source. We do not observe the same for Facebook, where still total low-credibility is comparable to the top-3 high-credibility domains. This shows that even though trustworthy information is more prevalent on both social media, the amount of disinformation content shared is still remarkable.

Finally, we investigate the relative popularity of low-credibility news websites on the two platforms, by computing Spearman's correlation coefficient of the websites ranked by their volumes. We find a significant positive correlation for low-credibility websites (R$= 0.65$, p-value$=1.14e-05$), indicating that the majority of unreliable sources is popular on both platforms.

\section{YouTube as a potential source of misinformation}
As an additional source of information about vaccines, we consider links to YouTube videos shared alongside Facebook and Twitter posts. Previous work \cite{donzelli2018misinformation} has shown that YouTube is used by both vaccine advocates and skeptics, and we aim to investigate the quality of videos shared on the two platforms. Overall, our dataset contains over 6 $k$ links to YouTube shared 21,407 times on Twitter and 132,553 on Facebook.

After extracting URLs pointing to YouTube from tweets and Facebook posts, we use their IDs to query the Youtube API and collect metadata available for such videos. We collected data for approximately 3 $k$ videos (published by 1.6 $k$ unique channels) shared on Twitter, and 3.2 $k$ videos (published by 1.5 $k$ unique channels) shared on Facebook. For approximately 300 videos (50 of which were present on both platforms) the API did not return any results, meaning these videos were removed from YouTube due to copyright or policy infringement. Such videos were shared over 800 times on Twitter and 6.5 $k$ times on Facebook. Following \cite{yang2020covid}, we argue that these videos might have contained suspicious and harmful content. However, we cannot confirm this hypothesis as they were deleted and are no longer available.

We manually inspected the top-20 videos based on the number of tweets and Facebook shares. On Twitter, these videos achieved a total of 5,511 retweets and 5,770,308 YouTube visualizations, while on Facebook they were shared  61,154 times and reached 11,521,158 visualizations. The number of YouTube views was extracted on March 18th. Interestingly, we find several popular videos, on both platforms, which are associated with anti-vax views and misleading information.

A relevant case is the 1st most shared video on Twitter (and 4th on Facebook) with the title "IL PARERE DEL PREMIO NOBEL LUC MONTAGNIER SULLA VACCINAZIONE ANTI-COVID [VIDEO IN ITALIANO]"\footnote{Translation: "The opinion of Nobel Prize Winner Luc Montagnier on COVID-19 vaccination". Available at \url{www.youtube.com/watch?v=kHGtn_vnrJ8}.}, with over 700 retweets, 4k Facebook shares, and 450k YouTube views. In this video, the Nobel prize winner Luc Montagnier refers to Moderna company as "sorcerer apprentices" stating that they only tested the vaccine on animals, and it's thus not possible to foresee the effects of the vaccine on humans. He also proposes alternative natural treatments against COVID-19 and states that vaccinating the whole population is not the solution.

We omit other examples for reasons of space, but we report that several other popular videos mention conspiracy theories behind the origin of the virus and/or the effects of vaccines as well as proposing alternative therapies and suggesting the audience not to get vaccinated.

The fact that through a simple manual evaluation we encounter almost a dozen of suspicious and, in some cases, explicitly harmful videos among most popular videos indicates that further investigation is required. Indeed, it appears that YouTube is a potential source of online misinformation about vaccines.


\begin{figure*}[!ht]
    \centering
    \includegraphics[width=\linewidth]{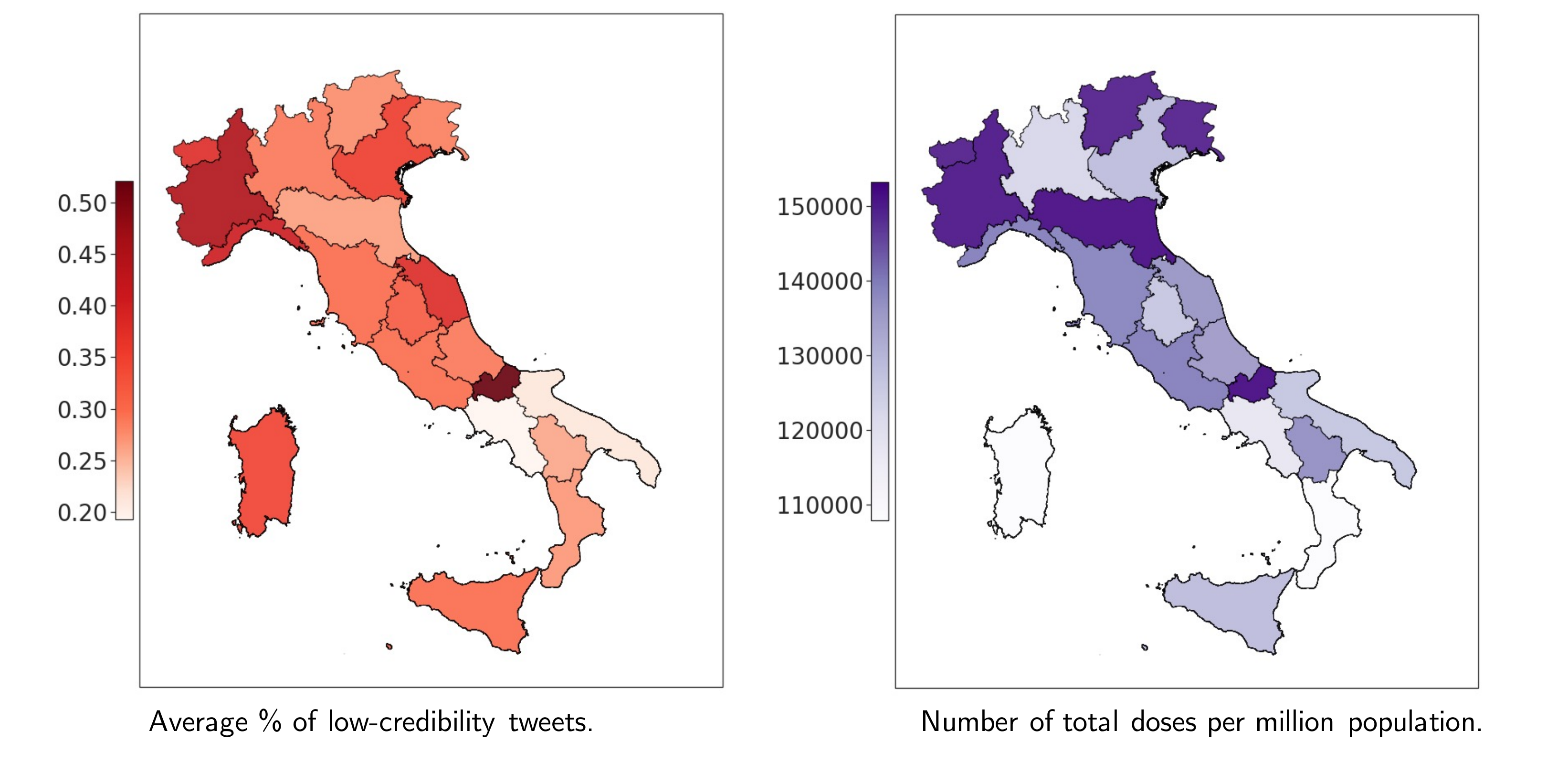}
    \caption{\textbf{Left.} Average fraction of low-credibility tweets shared by users, for each Italian region. \textbf{Right.}  Total amount of vaccine doses administered, per million population, in each Italian region.}
    \label{fig:geo}
\end{figure*}

\section{Geolocating Twitter conversations}
A goal of our project is to link online conversations with geographical details on the ongoing vaccination campaign, e.g. the number of doses administered in each Italian region.

To this aim, we attempt to geolocate Twitter users by using a naive string matching algorithm, i.e. checking whether they have a "location" field disclosed in their profile and matching it against a list of Italian municipalities, provinces, and regions\footnote{Taken from the Italian National Institute of Statistics and available at \url{https://www.istat.it}.}. In the case of multiple matches, we retain the longest one. We matched circa 16 $k$ unique locations and, among over 135 $k$ users putting a "location" in their profile, we accordingly geolocated 73 $k$ users to either an Italian municipality or region. These shared over 1.3 $M$ tweets. The number of accounts mapped to each Italian region is significantly positively correlated with the actual population (Pearson R$=0.89$, PVAL$< 0.001$). However, it is known that the Twitter sample of users might not be fully representative of the Italian population, and this is a limitation to analyses that infer demographics from Twitter~\cite{alessandra2017tweets}.

As an illustrative example, we show in Figure \ref{fig:geo} statistics on the amount of low-credibility information circulating on Twitter, and  the status of the vaccination campaign. Specifically, in the left panel, we show the average fraction of low-credibility tweets shared by users geolocated in each region; darker colors correspond to higher values. We note that on average, Italian users share low-credibility information around 0.20-0.50\% of the time. In the right panel, we show the total number of doses administered per million population, in each region. We can note that Lombardy is performing worse than most regions, even though it was the region most struck by the pandemic during the first wave.

These results are still preliminary, as the methodology presents several limitations and needs further assessment, e.g., how to handle multiple locations appearing in the "location" field of user profiles or when false places match with Italian municipalities with misleading names (e.g. "Paese" which translates as "village").

\section{Conclusions}
We present an ongoing project which monitors online conversations of Italian users around vaccines on Twitter and Facebook. We give full access to the data we are collecting, and we provide up-to-date results in an online interactive dashboard.
Preliminary analyses show that there is a non-negligible amount of low-credibility information circulating on both platforms, and they indicate YouTube as a potential source of misinformation about vaccines.
Our final goal is to understand the interplay between the public discourse on online social media and the vaccine roll-out campaign. In particular, we aim to investigate the impact of online sentiment (e.g. communities of pro and anti-vax) and misinformation about vaccines on vaccine uptake in Italy. We also aim to assess whether there are geographical socio-economic differences that shape both online conversations and the vaccination campaign.

\section{Acknowledgments}
This work has been partially supported by the PRIN grant HOPE (FP6, Italian Ministry of Education), and the EU H2020 research and innovation programme, COVID-19 call, under grant agreement No. 101016233 “PERISCOPE” (https://periscopeproject.eu/).

\bibliography{bib}

\end{document}